\documentclass[twocolumn,showpacs,preprintnumbers,amsmath,amssymb]{revtex4}

\usepackage{dcolumn}
\usepackage{color}
\usepackage{bm}
\usepackage{graphicx}
\usepackage{epsfig}
\usepackage{amsmath}
\usepackage{amsfonts}
\usepackage{amssymb}
\def \be{\begin{equation}}
\def \ee{\end{equation}}
\def \beq{\begin{equation}}
\def \eeq{\end{equation}}
\def \bea{\begin{eqnarray}}
\def \eea{\end{eqnarray}}

\def\bfeta{{\boldsymbol{\eta} }}

\def\br{{\bf r}}

\def\bc{{\bf c}}

\def\bk{{\bf k}}

\def\bi{{\bf i}}

\def\cH{{\cal H}}
\def\cS{{\cal S}}

\def\bfeta{{\boldsymbol{\eta}} }
\def\half{{\frac{1}{2}}}

\def\etal{{\it et.~al.~}}
\def\9{\rangle}
\def\6{\langle}

\def\be{\begin{equation}}
\def\ee{\end{equation}}
\def\bea{\begin{eqnarray}}
\def\eea{\end{eqnarray}}
\def\bq{{\bf q}}

\def\bk{{\bf k}}
\def\cH{{\cal H}}
\def\cS{{\cal S}}

\begin{document}
\title{Temperature Dependence Of  Cuprate Superconductors' Order Parameter}
\author{Alexander Mihlin and Assa Auerbach}
\affiliation{Physics Department, Technion, Haifa  32000, Israel}
\begin{abstract}
A model of charged hole-pair bosons, with long range Coulomb interactions and very weak interlayer coupling, is used to calculate the order parameter $\Phi$ of underdoped cuprates. Model parameters are extracted from experimental superfluid densities and plasma frequencies. The temperature dependence $\Phi(T)$ is characterized by a {\em 'trapezoidal'} shape. At low temperatures, it declines slowly due to harmonic phase fluctuations which are suppressed by anisotropic plasma gaps. Above the single layer Berezinski-Kosterlitz-Thouless (BKT) temperature, $\Phi(T)$ falls rapidly toward the three dimensional transition temperature. The theoretical curves are compared to $c$-axis superfluid density data by H. Kitano et al., (J. Low Temp. Phys. {\bf 117}, 1241 (1999)) and to the {\em transverse nodal velocity} measured by angular resolved photoemmission spectra on BSCCO samples by W.S. Lee {\em et  al.}, (Nature {\bf 450}, 81 (2007)), and by  A. Kanigel,  {\em et al.}, (Phys. Rev. Lett. {\bf 99}, 157001 (2007)).

\end{abstract}
\date{\today}

\maketitle
%
\section{Introduction} 

Unconventional superconductivity in cuprates is often measured by deviations from  Bardeen, Cooper and Schrieffer's (BCS) phenomenology \cite{bib:BCS}.  A case in point is the order parameter
\be
   \Phi (T) =  \sum_\bfeta  d(\bfeta)   \langle c^\dagger_{\br\uparrow} c^\dagger_{\br+\bfeta \downarrow}\rangle,
   \label{PiT}
\ee
where $d(\bfeta)$ is the intra-layer pairing function with $d$-wave symmetry, and uniformity is assumed
in suppressing the $\br$ dependence of $\Phi$.  In BCS theory, the order parameter is inextricably related to a {\em gap in the quasiparticle excitations}, whose maximal value is given by

\be
   \Delta_{BCS}(T)=\bar{V} \Phi(T).
   \label{DeltaBCS}
\ee 
where $\bar{V}$ is an interaction parameter. $\Delta_{BCS}$ is the pair breaking energy which sets the scale of the transition temperature $T_c$.
However, BCS theory is a mean field approximation which neglects all phase fluctuations.  

In underdoped cuprates, there is compelling evidence that $T_c$ is driven by phase fluctuations \cite{bib:Emery}.  Uemura's empirical scaling law  $T_c \propto\rho^{ab}_s(T=0)$ \cite{bib:Uemura} and the observation of a superfluid density jump in ultrathin underdoped  cuprate films \cite{BKT-IV, bib:2D-KT, bib:3D-KT, bib:Lemberger} are consistent with the behavior of a bosonic superfluid, captured by an effective $xy$  model. 

In this paper we calculate the temperature dependent order parameter of an effective Hamiltonian of charged lattice bosons (CLB). The CLB model incorporates essential ingredients of underdoped cuprates including extremely weak interlayer coupling, and long range Coulomb interactions. 

\begin{figure}
   \includegraphics[width=8cm]{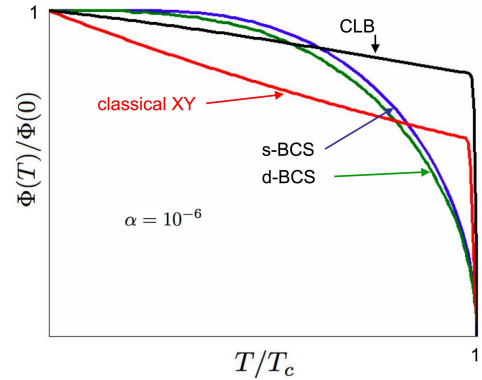}
   \caption{Temperature dependences of normalized superconducting order parameters. A {\em trapezoidal} shape is obtained for Charged Lattice Bosons (black color 
            online) model (Eq.\ref{Ham}) for anisotropy ratio $\alpha=10^{-6}$ and $\kappa=150$. We see that coulomb interactions suppress  thermal phase fluctuations 
            relative to the classical $xy$ model, Eq.(\ref{xy}), (red color online),  depicted for $\alpha=10^{-6}$. The rapid fall toward $T_c$ is calculated within  
            interlayer mean field theory (see text). BCS theory for $d$ and $s$-wave order parameters is depicted for 
            comparison, (green and blue colors online, respectively). }
   \label{Fig:BCS-all}
\end{figure}

Our main result is that  $\Phi(T)$ exhibits a  {\em trapezoidal shape} in the weak interlayer coupling limit, as depicted in Fig. \ref{Fig:BCS-all}. At low temperatures  $\Phi(T)$  decreases slowly due to effects of anisotropic plasma frequency gaps. The effects of long range charge interactions, however, do not drive the transition. The transition is driven by proliferation of vortex loops above the two dimensional Berezinskii-Kosterlitz-Thouless (BKT) \cite{bib:BKT} temperature $T_{BKT}$, where the order paramater falls rapidly toward $T_c$.

Phase fluctuation  theories  have been previously applied to cuprates, with special attention to  the intra-layer superfluid density  $\rho^{ab}_s(T)$ \cite{bib:Paramekanti,bib:Kwon,bib:Franz}. In order to explain the  linearly decreasing   temperature dependence, additional gapless (nodal)  fermionic excitations were argued to be essential \cite{Herbut}.

$\Phi(T)$,  however,  behaves differently than $\rho_s^{ab}(T)$. In the two dimensional limit, for example, $\Phi$ must vanish  at all $T>0$ by Mermin and Wagner theorem  \cite{bib:MW}, while $\rho_s^{ab}$ jumps to a finite value below $T_c$. Also, nodal fermions have a small effect on $\Phi(T)$. This is  demonstrated in Fig.\ref{Fig:BCS-all}, which shows $s$ and $d$ wave order parameters  behaving very similarly within BCS theory.
 
We propose  experimental probes for  the order parameter,  without relying on BCS theory and Eq. (\ref{DeltaBCS}).
At weak interlayer coupling,  we argue that $\Phi(T)$ should be proportional
to the  square root of the $c$-axis superfluid density.  

Angular Resolved Photoemmission Sprectroscopy (ARPES) finds a {\em ''pseudogap''} $\Delta_{pg}$  in the electronic spectrum, which persists
well above $T_c$ \cite{bib:Curty,bib:Hufner,bib:Panagopoulos99-2}.  Apparently, $\Delta_{pg}(T)$
is {\em not} proportional to $\Phi(T)$, (the latter of course vanishes at $T_c$), which violates Eq. (\ref{DeltaBCS}).  Pseudogap phenomena are often interpreted as {\em short range  pairing correlations} well above $T_c$.  

To  address  ARPES data, we employ a Boson-Fermion (BF) model which was derived from the Hubbard model \cite{AA} by contractor renormalization. The model describes the CLB system,  Andreev-coupled to  fermion quasiparticles which occupy small hole pockets (or 'arcs'). Similar BF models  were arrived at by  other approaches \cite{BFmodel,QED3,SubirBF}. Within our model,  $\Phi(T)$  is proportional  to the {\em transverse nodal velocity } $v_\perp(T)$. In Fig. \ref{Fig:arpes}, we find reasonable agreement between the theoretical curves and BSCCO ARPES data for $v_\perp(T)$ of Refs. \cite{bib:Lee,bib:Kanigel}. Further tests of the trapezoidal shape closer to $T_c$ would be desirable.
 
The paper is organized as follows: The CLB model is introduced in Section \ref{sec:CLB}. The order parameter is calculated within the harmonic phase fluctuations approximation to obtain the low temperature regime. In Section \ref{sec:IMFT}, the interlayer mean field theory is applied to compute the suppression of the order parameter near $T_c$. In Section \ref{sec:exp-par} we  relate the model parameters to experimental data for several commonly studied cuprates. Section \ref{sec:exp} compares the theory to  experiments,  using an effective Boson-Fermion model to interpret the ARPES data. We conclude with a brief summary and discussion.

In Appendix \ref{app:FitHPF}, we provide details of the analytical fit to the harmonic phase fluctuations result. In Appendix \ref{app:GC} we estimate the temperature
region near $T_c$ where three-dimensional critical fluctuations are important (Ginzburg criterion).

%
\section{Charged Lattice Bosons} 
\label{sec:CLB}

The $xy$ Hamiltonian is a lattice model of boson phase fluctuations,

\be
   \cH_{xy}=-\frac{J}{2}\left(\sum_{\br,\bfeta }\cos(\varphi_{\br}-\varphi_{\br+\bfeta})+\alpha\sum_{\br,\bc}\cos(\varphi_{\br}-\varphi_{\br+\bc})\right)   
   \label{xy}
\ee
where $\br$ resides on a layered tetragonal lattice. $\bfeta$ and $\bc$ are in-plane and interlayer nearest neighbor vectors of lengths $a$ and $c$ respectively. 
The lattice constant $a$ is in effect a coarse grained parameter chosen to be larger than the coherence length $\xi$.
$J$ is the bare intra-layer superfluid density, and $\alpha\ll1 $ is the anisotropy ratio.

The order parameter is defined by

\be
   \Phi(T) = \Phi_0\langle \cos (\varphi_\br)\rangle
   \label{OP}  
\ee
where $\Phi_0$ is the zero temperature value. The quantum  CLB  model is given by

\be
   \cH_{clb}=\cH_{xy}[\varphi]+\half \sum_{\br,\br'}V(\br-\br')n_{\br}n_{\br'}
   \label{Ham}
\ee
where $n_\br$ is the occupation number of a charge $2e$ boson on site $\br$, obeying the commutation relation,  

\be
   \left[ n_{\br},\varphi_{\br'}\right] =  i  \delta_{\br,\br'}
   \label{CR}
\ee

Long range Coulomb interactions $V(\br)$ are given by the Fourier components

\be
   V_\bq= \sum_{\br} e^{-i\bq\br} V(\br)= \frac{16 \pi e^2}{v \epsilon_b  q^2}.
   \label{Vq}
\ee
where $v\equiv a^2c$ is a unit cell volume and $\epsilon_b(\bq,\omega_\bq )$ is the effective dielectric function in the appropriate wave vector and frequency scale.

At low temperatures, we  can expand  the CLB action  to  quadratic  order and obtain the harmonic phase fluctuations (HPF) action,

\be
   \cS_{hpf}[\varphi]=\frac{1}{2}\hbar^2 T\sum\limits_{\bq n}\frac{\omega_n^2+\omega_p^2(\bq)}{V_\bq}\varphi_{\bq\omega_n}\varphi_{-\bq-\omega_n}
   \label{eq:s}
\ee
where $\omega_n=2\pi n T /\hbar $ are bosonic Matsubara frequencies. The plasmon dispersion, as derived by Kwon \etal  \cite{bib:Kwon}, is

\bea
   &&\omega_p^2(\bq) \equiv \frac{\omega_{ab}^2q_{ab}^2+\omega_c^2 q_c^2}{q^2}\nonumber\\
   &&\omega^2_{ab} = \frac{16\pi e^2 J }{\epsilon_b \hbar^2 c}\nonumber\\
   &&\omega^2_{c} = \frac{16\pi e^2 c\alpha J}{\epsilon_b \hbar^2 a^2},
   \label{Plasma}
\eea
where $q_{ab}$ and $q_c$ are the planar and $c$-axis wave-vectors respectively.

The HPF order parameter is given by

\be
   \Phi_{hpf}(T) = \Phi_0 e^{-\frac{1}{2}\langle\varphi_\br^2\rangle}
   \label{M_hpf}
\ee
where the local phase fluctuations are given by,
\bea
   \langle\varphi_{\br}^2\rangle &=& \frac{1}{Z}\int{\cal D}\varphi ~ \varphi_{\br}^2 e^{-\cS_{hpf}[\varphi]}\nonumber\\
   &=& v\int\!{d^3 q\over(2\pi)^3}\frac{V_\bq}{\hbar\omega_p(\bq)}\left(\frac{\sinh( \hbar\omega_p(\bq)/T)}
   {\cosh( \hbar\omega_p(\bq)/T)-1}\right) \nonumber\\  
   \label{phi-cyl}
\eea

At extremely low temperatures, $T\ll\hbar\omega_c$, all thermal phase fluctuations are frozen out. However, as we shall show in Section \ref{sec:exp}, the experimentally interesting regime of large anisotropy, has a wide separation of plasma energy scales, such that

\be
   \hbar\omega_c \ll T_{BKT}\sim T_c   \ll \hbar\omega_{ab}
\ee

For our regime, we fit  Eq. (\ref{phi-cyl}) by the analytical approximation  (see Appendix \ref{app:FitHPF}),

\be
   \langle\varphi^2\rangle_{T,\alpha}=\left({ T \over J}\right)\left( a_1 -a_2 |\ln(\alpha)|\right)  e^{-a_3 \hbar \sqrt{\omega_{ab}\omega_c}/T}
   \label{phi-low-T}
\ee

For the simplified case of $a=c$, the coefficients are given by:

\bea
   a_1 & \approx & 0.045,~~~a_2= - 0.013\nonumber\\
   a_3 &\approx & 0.35 
   \label{phi-num-coef}
\eea

Thus, expression (\ref{M_hpf}) reduces to the classical result of Hikami and Tsuneto (HT) \cite{bib:Hikami}, (shown later in Eq. (\ref{PL})) in the limit $T\gg a_3\hbar\sqrt{\omega_{ab}\omega_c}$. In the experimentally relevant regime, $\Phi_{hpf}$ decreases significantly slower than the classical model, as demonstrated in Fig. \ref{Fig:BCS-all}.
%
\section{Interlayer Mean Field Theory}
\label{sec:IMFT}

   The HPF action (\ref{eq:s}) cannot describe the order parameter near $T_c$ since it does not include  vortex excitations. 
In the narrow regime of $T_{BKT}  \le T\le T_c$ proliferation of  widely separated two dimensional vortex pairs dramatically reduces 
the order parameter. 

   For anisotropies  of order  $\alpha \sim  10^{-4} - 10^{-6}$,  a straighforward numerical calculation of Eq. (\ref{xy})
is encumbered by finite size limitations.  Instead, we employ the interlayer mean field theory (IMFT) \cite{IMFT}, described  by a single
layer hamiltonian in an effective field $h$:

\bea
   {\cal H}_{imft}(h) &=& \cH_{2d}(h)+\frac{h^2}{2 \alpha J} \nonumber\\
   \cH_{2d}(h) &=& -J\sum^{2D}_{\br\bfeta}\cos(\varphi_\br-\varphi_{\br+\bfeta}) - 2h \sum^{2D}_\br \cos(\varphi_\br) \nonumber\\
   \label{H2d}
\eea

Variational detrminition of $h$ yields the IMFT equation

\be 
   h = {2\alpha J}\langle\cos\varphi_\br\rangle = 2\alpha J\Phi_{2d}(T,h)
   \label{m_MF}
\ee 
where the magnetization of a single two dimensional layer, $\Phi_{2d}(T,h)$, is, in principle, the {\em exact} field dependent order parameter of the single layer CLB model. Solving Eq. (\ref{m_MF}) for $h(T)$, yields the three dimensional temperature dependent order parameter 

\be
  \Phi_{imft}(T) = \Phi_{2d}(T,h(T))  .
   \label{m_imft}
\ee

The transition temperature $T_c$ is given by

\be
   T_c=\mathop{\min}\limits_T\left\{{T;\Phi_{imft}(T)=0}\right\}
\ee

Solution of eq. (\ref{m_MF}) for samll anisotropies requires precise determination of $\Phi_{2d}(h,T)$ for very weak fields $h$ near $T_c$. This is obtained by using the asymptotic critical properties of the order parameter near $T_{BKT}$, which is not far from $T_c$ in the small $\alpha$ limit.
%
\subsection{BKT critical properties} 
\label{sec:clb}

   The two dimensional classical $xy$ model undergoes a BKT transition  \cite{bib:BKT} at $T_{BKT} \approx 0.89 J$  \cite{bib:Tobochnik,bib:T_KT}. Vortex pair proliferation changes the phase correlation temperature dependence from power law to exponential decay,

\bea
   \langle \cos(\varphi_\br- \varphi_0)\rangle  &\sim & \begin{array}{ll}
   r^{-\eta(T)}& T<T_{BKT}\\
   e^{-r/\xi(T)} & T> T_{BKT}\end{array}
   \label{2Dcorr}
\eea
where at low temperatures,

\be
   \eta \simeq {T\over 2\pi J}, ~~~T\ll T_{BKT}
\ee 

Above $T_{BKT}$ the correlation length diverges  as

\bea
   \xi_{2d} &\propto &  \exp\left(  \beta/\sqrt{t}\right)\nonumber\\
	 \chi_{2d}(t) &=& { B_\chi\over J } \exp\left( \nu \beta/\sqrt{t}\right)\nonumber\\
   t &\equiv &(T-T_{BKT})/T_{BKT} \nonumber\\
   \beta&=& 3/2,~~~\nu=7/4
   \label{xy-crit}
\eea

where the exponents $\beta$ and $\nu$ were derived by Kosterlitz \cite{bib:Kosterlitz}. 

\begin{figure}
   \includegraphics[width=8cm]{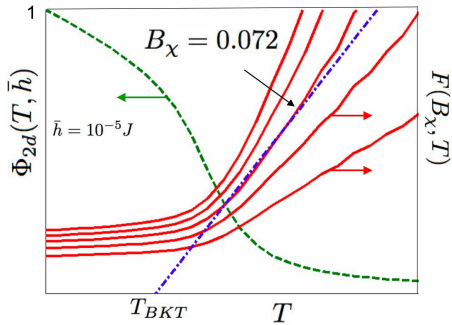}
   \caption{Determination of $B_\chi$ from Monte-Carlo data. The different fitting functions $F(B_\chi,T)$ (solid lines, red color online),  are defined in 
            Eq.(\ref{fit}). The curve with parameter value $B_\chi=0.072$ is chosen as the best fit to $\left(T-T_{BKT}\right)/T_{BKT}$. The dashed (gren color online) line is the two dimensional order parameter in the presence of
            an ordering field $\bar{h}$.}
   \label{fig:fit}
\end{figure}

In order to match the transition region to the low temperature HPF order parameter, we need to determine the
non-universal amplitude $B_\chi$ of  $\chi_{2d}(T)$.  $B_\chi$   was determined numerically. 
We evaluated 
$\Phi_{2d}(h,t)$ by a Monte-Carlo simulation with Hamiltonian (\ref{H2d}). Good convergence was achieved with $10^9$ spin tilts per $(T,h)$ point, sampling every $10^5$ tilts and averaging over the last 5000 configurations.  We define a fitting function
\be
F(B_\chi,T) = \left( { \nu\beta\over \ln\left( J \Phi_{2d}(h,T)/(h B_\chi ) \right)} \right)^2
\label{fit}
\ee
 The fitting procedure  which is depicted in Fig. \ref{fig:fit}, yields

\be
   B_\chi \simeq 0.072.
   \label{B_chi}
\ee

   For finite interlayer coupling, the classical $xy$ model orders at $T_c (\alpha)> T_{BKT}$. Hikami and Tsuneto \cite{bib:Hikami} evaluated the order parameter for small $\alpha\ll1$, and obtained

\be
   \Phi_{cl}(T) = \Phi_0 \alpha^{\eta/(4-2\eta)}\simeq  \Phi_0 e^{-{T\over 8\pi J}| \ln\alpha|}.
   \label{PL}
\ee

In Fig. \ref{Fig:BCS-all}, $\Phi_{cl}(T)$ of Eq. (\ref{PL}) is plotted in comparison to the CLB model. The classical model decreases much faster since it does not contain the plasma gaps in the thermal phase fluctuations.

The IMFT equation for $T_c$ is

\be
   2\alpha J\chi_{2d}(T_c)=1.
   \label{Tc}
\ee 

Using Eq. (\ref{xy-crit}) for $\chi_{2d}(T)$ and the value (\ref{B_chi}) for $B_\chi$, the shift of  $T_c$ is

\be     
   T_c-T_{BKT} \sim \left(\frac{\beta\nu}{\ln(2B_\chi \alpha)}\right)^2 T_{BKT},
   \label{DeltaTc}
\ee
The IMFT is consistent with the renormalization group analysis of Hikami and Tsuneto \cite{bib:Hikami}. We note, however, that a large vortex core energy can  increase the shift of $T_c$ above the value given by Eq. (\ref{DeltaTc}) \cite{comm-core,bib:Giamarchi}.

The critical field-exponent was derived  by Kosterlitz \cite{bib:Kosterlitz} 

\be
   \Phi_{2d}(T_{BKT})\propto h^{1/\delta}, ~~~ \delta=15.
   \label{mh}
\ee

Combining this result with the IMFT equation (\ref{m_MF})  yields

\be
   \Phi(T_{BKT}) \propto \alpha^{1/(\delta-1)} = \alpha^{1/14} .
   \label{eq:m-alpha}
\ee   

Thus, by Eqs. (\ref{DeltaTc}) and (\ref{eq:m-alpha}), the order parameter drops rapidly  between $T_{BKT}$ and $T_c$, with an average slope of $d\Phi(T)/dT \sim - |\ln(\alpha)|^2$.
%
\subsection{Matching at the crossover} 

   In the crossover region, $\Phi_{2d}$ is given by the {\em harmonic mean} of the temperature and field dependent singularities at $T_{BKT}$.

\be
   \Phi_{2d}(T,h) = \Phi_{hpf}(T) \left( {1\over h \chi_{2d}(T)} + \left({h_0\over h}\right)^{1\over \delta} \right)^{-1}
   \label{eq:m-form}
\ee   

Eq. (\ref{eq:m-form}) correctly captures the singularities of the variables $(t,h)$ at  the BKT transition. $h_0$  is chosen to match the order parameter smoothly at $T_{BKT}$,

\be
   h_0 = 2 \alpha J \Phi_{hpf}(T_{BKT}),
\ee 

IMFT, as a mean field theory cannot properly capture  {\em three dimensional} critical exponents of the $xy$ model. Nevertheless, as shown in  Appendix \ref{app:GC}, the critical regime, by Ginzburg's criterion is limited to

\be
     T_c - T   < T_{BKT}/ |\ln \alpha |^4,
\ee
which is difficult to resolve experimentally, in the systems of interest.
%
\subsection{Fermionic excitations}

The CLB model ignores effects of fermionic particle-hole excitations, which are clearly observed in ARPES and tunneling. In underdoped cuprates, most of their spectral weight is associated with wave vectors around the antinodes, ($(\pi,0), (0,\pi)$), with energies at the pseudogap scale $\Delta_{pg}\gg T_c $. Contribution of these excitations to depletion of the order parameter temperature is of order $T/\Delta_{pg}\ll1$.

Nevertheless, one might worry that {\em low energy} (nodal) excitations  might play an important role. This has been shown to be the case for the  temperature dependence of the  superfluid density $\rho^{ab}_s(T)$ \cite{WenLee,bib:Franz,bib:Paramekanti}). 

However, nodal excitations are weakly coupled to the order parameter. Consider, for example,  the BCS gap equation,

\be
   {1\over \lambda} = \sum_{\bk}  {|d(\bk)|^2 \over E_{\bk}(\Delta(T)) } \tanh\left( E_{\bk}(\Delta(T)) /  T\right) ,
\ee
where

\be
   E_\bk=\sqrt{(\epsilon_\bk-\mu)^2 + |d(\bk) \Delta(T) |^2},
   \label{Ek-SC}
\ee
and $\lambda$ is the BCS coupling constant. The pair wave function factor $|d(\bk)|^2$ vanishes on the nodal lines $\bk=(\pm k, k)$. This suppresses contributions from the nodal regions to the thermal depletion of the gap. As a result, $s$-wave and  $d$-wave  order parameters have very similar temperature dependence as  shown  by Won and Maki \cite{bib:Maki} and depicted in Fig. \ref{Fig:BCS-all}. Although here we do not appeal to BCS theory, this  observation depends only on the weak coupling between nodal fermions and  the order parameter, imposed by the pair wave function symmetry.

\begin{table*}
   \begin{center}
      \begin{tabular}{|l|c|c|c|c|c|c|c|c|c|c|}
         \hline
         Compound & $a$ [A] & $c$ [A] & $T_c$ [K] & $\hbar\omega_{ab}$ [eV] & $\hbar\omega_c$ [meV] & $\Omega$ [meV] & $\lambda_{ab}$ [$\mu$m]
                  & $\lambda_c$ [$\mu$m] & $\alpha$ [$10^{-4}$] & References \\
         \hline
         $\rm{YBa_2Cu_3O_{7-\delta}}$         & 3.8 & 5.8 & 89  & 1.5-2.5   & 5.6-13.6 & 36.1-72.3 & 0.14-0.28 & 1.26-7.17 & 50-5  &     
            \cite{bib:Farnan,bib:Qiu,bib:Kojima,bib:Panagopoulos98}\\
         $\rm{Bi_2Sr_2CaCu_2O_{8+\delta}}$    & 5.4 & 7.7 & 92  & 0.94-1.84 & 0.23-1.4 & 5.6-19.5  & 0.2       & 110       & 0.016 & 
            \cite{bib:Colson,bib:Uchida,bib:Motohashi,bib:Gaifullin}\\
         $\rm{La_{2-\delta}Sr_{\delta}CuO_4}$ & 3.8 & 6.6 & 40  & 0.3-3     & 3.7-11.2 & 13-75.4 & 0.19-0.28 & 2-8.5     & 30-3  &
            \cite{bib:Qiu,bib:Dordevic03,bib:Dordevic05,bib:Ortolani,bib:Panagopoulos99-1,bib:Panagopoulos99-2}\\
         $\rm{Tl_2Ba_2CaCu_2O_{8+\delta}}$    & 3.9 & 7.4 & 108 & 1.5       & 1.2-2.6  & 17.4-26.8  & 0.17-0.33 & 2.5-8.4   & 13-4  &
            \cite{bib:Tominari,bib:Thorsmolle,bib:Wang}\\
         \hline
      \end{tabular}
	 \end{center}
   \caption{Typical planar lattice constants, $a$, mean interplane distances, $c$, critical temperatures, $T_c$, planar and interplane plasma frequencies, 
            $\omega_{ab}$, $\omega_c$, energy scales $\Omega(\alpha,a,c)$ of Eq. (\ref{phi-low-T}), magnetic field penetration depths, $\lambda_{ab}$, $\lambda_c$ 
            and anisotropy factors, $\alpha$, at zero temperature. All quantities except for $\Omega(\alpha,a,c)$ and $\alpha$ were obtained experimentally, while 
            $\Omega$ and $\alpha$ were obtained via Eqs. (\ref{Omega}) and (\ref{alpha}) respectively. Some quantities depend on doping (e.g. 
            $\lambda_{ab},\lambda_c$ are diminished with doping) and values for each compound correspond to similar dopings.}
   \label{tab:prop}
\end{table*}
%
\section{Experimental Parameters}
\label{sec:exp-par}

The cuprates exhibit very large anisotropy between in-plane and interlayer Josephson couplings $J_c$ and $J_{ab}$, which can be experimentally determined by the in-plane and interlayer zero temperature London penetration depths $\lambda^0_{ab}$ and $\lambda^0_{c}$,

\bea
   \lambda^0_{ab}=\left(\frac{16\pi e^2}{\hbar^2c^2d}J_{ab}\right)^{-\half}\nonumber\\
   \lambda^0_{c}=\left(\frac{16\pi e^2d}{\hbar^2c^2a^2}J_{c}\right)^{-\half},
   \label{lambda}
\eea
where $d$ and $a$ are effective lattice constants, $e$ is the electron charge and $c$ the speed of light. The anisotropy ratio for cuprates is in the range,

\be
   \alpha \equiv \frac{J_c}{J_{ab}} = \left({\lambda^0_{ab}a\over\lambda^0_c d}\right)^2\sim 10^{-6}-10^{-3}.
   \label{alpha}
\ee

Our phenomenological assignment of $J_{ab}$ and $J_c$, neglects quantum corrections which become sizable near the critical doping toward  the insulating phase. An alternative measure of $J_{ab}$ and $J_c$ is given by relations (\ref{Plasma}) and the experimental measurements of $\omega_{ab}$ and $\omega_c$ by optical and microwave conductivities (cf. Refs. \cite{bib:Farnan,bib:Motohashi,bib:Colson,bib:Dordevic05,bib:Ortolani,bib:Thorsmolle,bib:Qiu,bib:Kojima,
bib:Uchida,bib:Dordevic03,bib:Tominari}). Thus, the anisotropy parameter, $\alpha$, of Eq. (\ref{alpha}) can be determined.
 
Table \ref{tab:prop} contains typical experimental values of relevant quantities at zero temperature (except for $\Omega$ and $\alpha$ which were determined via Eqs. (\ref{Omega}) and (\ref{alpha}) respectively).
In YBCO, BSCCO and TBCCO,  the interplane distance  $c$ is taken as the mean value.
%
\section{Experimental Probes of $\Phi(T)$}
\label{sec:exp}

\begin{figure}
   \includegraphics[width=8cm]{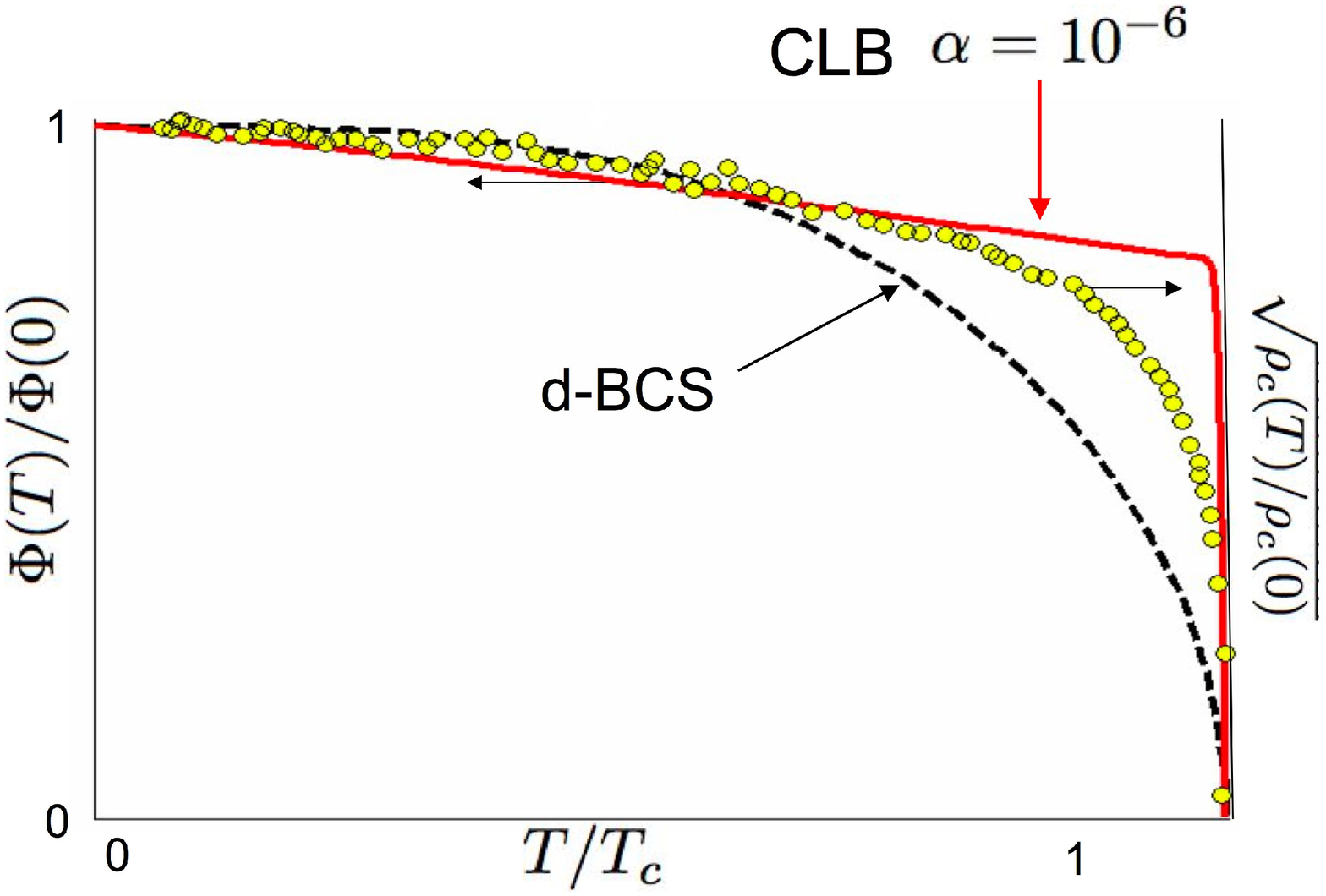}
   \caption{Comparison of CLB order parameter to square root of $c$-axis superfluid density from Ref. \cite{bib:Kitano}. Model parameters are $\alpha=10^{-6}$, 
            $c/a=0.5$ and $\kappa\equiv\omega_{ab}/J=150$. Data was taken on BSCCO with $T_c=87$ K. Dashed line is d-wave BCS energy gap,  given   for comparison.}
   \label{Fig:rho-c}
\end{figure}

In cuprates, the BCS relation, (\ref{DeltaBCS}), does not hold, since the maximal gap $\Delta_{pg}$ is weakly temperature dependent \cite{bib:Hufner,bib:Timusk}, while $\Phi(T)$  vanishes at $T_c$. Here we propose experimental probes to measure $\Phi(T)/\Phi(0)$.
%
\subsection{$c$-axis superfluid density}
Since the zero temperature interlayer pair tunneling is  weak,  the layered system can be treated as a one dimensional array of Josephson junctions. Within a  variational approximation,  the order parameter can be extracted from the temperature dependence of the $c$-axis superfluid density, 

\be
   \rho_s^c(T) = \rho_s^c(0)  |\Phi(T)|^2  .
\ee

Indeed, as seen in Fig. \ref{Fig:rho-c}, agreement between theoretical curves $\Phi(T)$ and values extracted from 
electrodynamical data of BSCCO \cite{bib:Kitano} are quite good, except near the transition.  
%
\subsection{ARPES}
In $d$-wave BCS theory  the  quasiparticle spectrum is given by Eq. (\ref{Ek-SC}).
Above $T_c$, $\Delta_{BCS}=0$, and the full Fermi surface should be detected as zero energy crossings of the  ARPES quasiparticle peaks.
However,  in underdoped cuprates  as temperature is raised above $T_c$, only finite Fermi arcs appear around the nodal directions. The gap  in the anti-nodal directions $\Delta_{pg}$ survives to much higher temperatures \cite{bib:Hufner,bib:Timusk}. In contrast to $\Delta_{pg}$,  the transverse nodal velocity $v_\perp$ vanishes abruptly at $T_c$
\cite{bib:Lee,bib:Kanigel}. Below $T_c$, $v_\perp$ introduces  a singularity $|k_\perp|$  in the electronic propagator, which translates to an infinite correlation length in real space.

A microscopic connection between $v_\perp(T)$ and $\Phi(T)$  can be provided by an effective Boson-Fermion hamiltonian with small hole pockets, described below.

\begin{figure}
   \includegraphics[width=8cm]{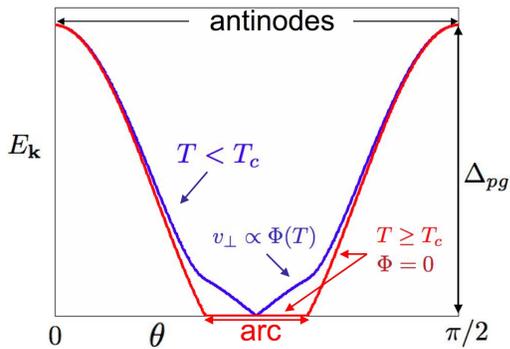}
   \caption{Boson-Fermion model for the transverse quasiparticle excitations below and above $T_c$. $\theta$ is the azimuthal coordinate transverse to the nodal 
            direction. Above $T_c$ (red color online),  vanishing of $E_\bk$ on a finite 'arc' reflects the inner edge of the hole pocket. The pseudogap $\Delta_{pg}$ 
            is the hole fermions energy at the antinodal wavevectors  $(\pi,0),(0,\pi)$, which has no direct bearing on the superconducting properties. Below  $T_c$  
            (blue color online), the Andreev coupling of  hole fermions to  hole-pair bosons yields a $d$-wave gap with a node at $\theta=0$. The transverse nodal 
            velocity $v_\perp(T)$ is a direct measurement of  $\Phi(T)$. The break in the curve at the arc edge is consistent with 'two gaps' phenomenology \cite{twogaps}.}
   \label{Fig:v-perp}
\end{figure}
%
\subsection{Boson-Fermion theory}

The Boson-Fermion (BF) model,  which arises by a contractor renormalization of the square lattice Hubbard model \cite{AA},  describes  
spin half fermion  holes $f_{\bk,s}$ of charge $e$, coupled to the CLB as

\bea
   \cH_{bf} &=&  \cH_{clb}  + \sum_{\bk,s} (\epsilon^h_\bk-\mu) f^\dagger_{\bk s}f_{\bk s} \nonumber\\
   &&+ g \sum_{\br,\br'} e^{i\varphi_\br}  d(\br-\br') f_{\br,\uparrow} f_{\br',\downarrow} + \mbox{h.c.}.
   \label{BFmodel}
\eea 

The last Andreev coupling term,  describes disintegration of hole pairs into  single spin-half hole fermions.
In our version of the BF model, the   fermion and boson densities, measured  with respect to  half filling, obey
\be
   n_h + 2 n_b = x,
\ee
where $x$ is the total concentration of doped holes. The hole dispersion $\epsilon_\bk$  has minima  near  $(\pm \pi/2,\pm \pi/2)$, and therefore occupy four {\em small} pockets of area fraction $n_h/2 $. Above $T_c$, the small  wave vector  sides of the pockets appear as the celebrated Fermi 'arcs' \cite{comm-dark}.The pseudogap is given by the quasiparticle excitation energy at the anti-nodal wavevectors
\be
   \Delta_{pg} = \epsilon_{(\pi,0)}-\mu .
\ee

In the superconducting phase $\Phi(T)=\langle \cos(\varphi)\rangle$. The hole fermions acquire the Dirac cone  dispersion
near the nodes:
\bea
   E_\bk = \pm \sqrt{ (v_F (k_\parallel-k_F))^2 +   (2g \Phi (T)  k_\perp)^2 },
   \label{Ebf}
\eea
that is depicted in Fig. \ref{Fig:v-perp}. Thus,   the transverse velocity directly measures  the  order parameter,
\be
   v_\perp (T) = 2g \Phi(T).
\ee
In the underdoped regime, the transverse velocity is smaller than the  pseudogap scale  $\Delta_{pg}d'(\bk)$.
This is seen as a 'break' in $E_\bk$ at the Fermi arcs angles, as shown in Fig. \ref{Fig:v-perp}.
Such behavior has been observed in ARPES \cite{twogaps} and found consistent with  a 'two gaps' phenomenology.

\begin{figure}
   \includegraphics[width=8cm]{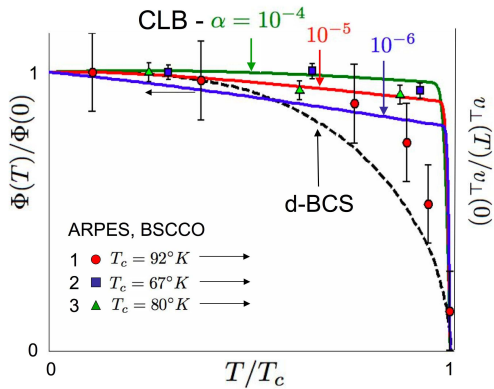}
   \caption{Comparison of CLB  to transverse nodal velocity, measured on different samples by ARPES. BSSCO samples with  $T_c$ noted in the figure. Experimental data, 
            including error bars, are (1) from Ref.\cite{bib:Lee}, (2) and (3) from Ref. \cite{bib:Kanigel}. The zero temperature normalization  is chosen by the 
            lowest temperature data points. Theoretical curves for several values of $\alpha$ are drawn. using $c/a=0.5$ and $\kappa\equiv\omega_{ab}/J=150$.  
            Dashed line is d-wave BCS energy gap,  given   for comparison.}
   \label{Fig:arpes}
\end{figure}

In Fig. \ref{Fig:arpes} we compare the CLB order parameter to the transverse nodal velocities measured on three samples of BSSCO by two groups \cite{bib:Lee,bib:Kanigel}. The agreement is reasonable, although the sharp break in the curves is not clearly confirmed. A comparison to the $d$-wave BCS expression shows a systematic trend of all the data being higher than BCS theory would predict.

%
\section{Discussion}

This paper calculated the order parameter of cuprates using  a bosonic model of  hole pairs. The  model includes crucial features of  layered cuprates: long range Coulomb interactions and very small anisotropy ratio. It ignores effects of fermionic particle hole excitations which are argued to be small for $\Phi(T)$.
The calculation predicts a trapezoidal temperature dependence in the small $\alpha$ limit, which is  distinct from both BCS  theory and the classical $xy$ model. The theoretical curves are compared  to data where the order parameter is extracted by additional theoretical assumptions: the $c$-axis superfluid  density (using a variational argument) and the transverse nodal velocity (using a BF model  of small hole pockets). We have selectively chosen data of BSCCO where $\alpha=10^{-6}$, and the 'trapezoidal' temperature dependence is most pronounced. In other cuprates, with larger values of $\alpha$, and larger vortex core energies \cite{comm-core}  the shift $Tc-T_{BKT}$ is larger, and the curve should be more rounded (less trapezoidal) and similar to the BCS curve.

Additional probes to $\Phi(T)$ could be devised. The critical current  of  a $c$-axis Josephson junctions with a higher $T_c$ material might be investigated.
The transverse nodal velocity, which we have related to $\Phi$ by the BF theory,  determines the low energy tunneling spectra and Raman scattering \cite{Raman}. In addition, it has been theoretically related to the linear slope of  the superfluid density    $d\rho_s^{ab}/dT$ \cite{WenLee}, and to thermal conductivity.
 
Further comparisons to experiments are warranted. Their success or failure may shed light on the applicability of the quantum lattice bosons  description of cuprates both below and above $T_c$. This would help us resolve some of the other mysteries of the pseudogap phase.
%
\section{Acknowledgements}

 We thank Ehud Altman, Thierry Giamarchi,  Amit Kanigel, Amit Keren and  Christos Panagopoulous, for useful advice and information. AA acknowledges support from the Israel Science Foundation, and is grateful for the hospitality of Aspen Center for Physics where some of the ideas were conceived.
%
\appendix
%
\section{Fitting Phase Fluctuations}
\label{app:FitHPF}

We define $q^2 \equiv q_c^2+q_{ab}^2$, $\eta^2\equiv\left(\omega_c/\omega_{ab}\right)^2=\alpha\gamma^2$ and $\gamma\equiv c/a$. For ease of numerical integration Eq. (\ref{phi-cyl}) may be simplified as follows

\bea
   \langle\varphi_{l\bi}^2\rangle &=& {1\over Z}\int{\cal D}\varphi ~ \varphi_{l\bi}^2 e^{-\cS^{(2)}[\varphi]}\\
   &=& v\int\!{d^3 q\over(2\pi)^3}\frac{V_\bq}{\hbar\omega_p(\bq)}\left(\frac{\sinh(\beta\hbar\omega_p(\bq))}
   {\cosh(\beta\hbar\omega_p(\bq))-1}\right)\nonumber\\
   &\approx& \frac{\gamma\hbar\omega_{ab}}{2\pi^2J}\int\limits^{\gamma\pi}_0dz\int\limits^{\pi}_0dr
   \frac{r}{\left(z^2+r^2\right)\varepsilon\left(\eta,\frac{z}{r}\right)}\nonumber\\
   &&\times\frac{\sinh\left[\varepsilon\left(\eta,\frac{z}{r}\right)/T\right]}{\cosh\left[\varepsilon\left(\eta,\frac{z}{r}\right)/T\right]-1},
   \label{phi-cyl1}
\eea
where the last expression was obtained in cylindrical coordinates. The dispersion is thus parametrized by

\be 
   \varepsilon\left(\eta,\frac{z}{r}\right) \equiv \hbar\omega_{ab}\sqrt{\frac{1+\eta^2\left(z/r\right)^2}{1+\left(z/r\right)^2}}.
\ee

   At extremely low temperatures, $T\ll\hbar\omega_c$,  all thermal phase fluctuations are frozen out. However, due to the large anisotropy, and poor screening, there is a wide separation of energy scales between the interplane plasma gap, $\hbar\omega_c$, and the planar gap, $\hbar\omega_{ab}$ and it turns out that 

\be
   \hbar\omega_c \ll T_c\sim J  \ll \hbar\omega_{ab}.
\ee

At low temperatures, the integral in Eq. (\ref{phi-cyl}) may be parametrized as

\be
   \langle\varphi^2\rangle \approx \frac{AT}{J}e^{\Omega/T},
   \label{app-phi-low-T}
\ee
where the energy scale, $\Omega(\alpha,\gamma)$, and the coefficient $A(\alpha,\gamma)$ may be parametrized by

\be
   \Omega(\alpha,\gamma) \approx \hbar\left(\frac{0.09}{\sqrt{\gamma}}+0.26\sqrt{\gamma}\right)\sqrt{\omega_{ab}\omega_c},
   \label{Omega}
\ee
and
\bea
   &&A(\alpha,\gamma) \approx A_1(\gamma)-A_2(\gamma)\ln(\alpha) \nonumber\\
   &&A_1(\gamma) \approx 0.029\gamma+0.016\gamma^2 \nonumber\\
   &&A_2(\gamma) \approx 0.24\gamma-0.11\gamma^2.
   \label{app-phi-num-coef}
\eea
The low temperature magnetization, $\Phi_{hpf}\left(T\right)$, is given by

\be
   \Phi_{hpf}(T) \propto e^{-\frac{1}{2}\langle\varphi^2\rangle}=C\left(T\right)\alpha^{\frac{A_2T}{2J}\exp(\Omega/T)},
   \label{m_ph}
\ee
where the coefficient $C\left(T\right)$ is given by
\be
   C\left(T\right)=e^{-\frac{A_1T}{2J}\exp(\Omega/T)}.
   \label{m_ph_coef}
\ee

Notably, $\gamma$ is of order unity and the energy scale, $\Omega(\alpha,\gamma)$, in Eq. (\ref{app-phi-low-T}) is proportional to the geometric average of the interplane and planar plasma energies.
%
\section{Ginzburg's Criterion for Interlayer Mean Field Theory}
\label{app:GC}

One would like to know, in which regime can we trust the IMFT near the transition temperature. Here we estimate the critical region using the standard Ginzburg Criterion. At small $\alpha$,  we see that the magnetization only varies rapidly below $T_c$, in the narrow region of width $\Delta T_c$ given by Eq. (\ref{DeltaTc}). Within that region, $\Phi_{imft}(T)$ drops from  $\Phi_{hpf}(T_{KT})$, as given by the harmonic phase fluctuations (\ref{m_ph}), to zero at $T_c$, with a mean field behavior,

\be
   \Phi_{imft} \sim  \Phi_{hpf}(T_{KT}) \left({|T-T_c|\over\Delta T_c}\right)^\beta,~~~~\beta=\half.
   \label{exp:beta}
\ee

Ginzburg's criterion \cite{Ginzburg,CL}, estimates the temperature region below $T_c$, where critical 3D fluctuations become important and IMFT breaks down. This is where order parameter fluctuations averaged over a correlation volume of size $V_\xi = \xi^2_{ab}\xi_c$ exceed their average, i.e.

\be
   \langle(\Delta \Phi)^2\rangle_{V_\xi} = {S(\bq=0,T)\over V_\xi} = {c\over\xi_c(T)} \ge \Phi_{imft}^2(T) .
\ee

Using the mean field estimation of  $\xi_c \sim c(|T-T_c|/T_c)^{-\half}$ and Eq. (\ref{exp:beta}), the critical regime is given by

\be
   |T-T_c| \le \Delta T_c^2/T_c \ll \Delta T_c,
\ee
which is much smaller than the already narrow region of $\Delta T_c$, where 2D vortex pair fluctuations suppress the order parameter. In summary, for layered systems with large anisotropy, IMFT theory holds up to temperatures very close to $T_c$.
%


\begin{thebibliography}{100}

\bibitem{bib:BCS}
   J. Bardeen, L. N. Cooper, J. R. Schrieffer, 
   Phys. Rev., {\bf 108}, 1175 (1957).
\bibitem{bib:Emery}
   V.J. Emery, S.A. Kivelson,
   Nature, {\bf 374}, 434 (1995).
\bibitem{bib:Uemura}
   Y.J. Uemura, G.M. Luke, B.J. Sternlieb, J.H. Brewer, J.F. Carolan, W.N. Hardy, R. Kadono, J.R. Kempton, R.F. Kiefl, S.R. Kreitzman, P. Mulhern, 
   T.M. Riseman, D.Ll. Williams, B.X. Yang, S. Uchida, H. Takagi, J. Gopalakrishnan, A.W. Sleight, M.A. Subramanian, C.L. Chien, M.Z. Cieplak, G. Xiao,
   V.Y. Lee, B.W. Statt, C.E. Stronach, W.J. Kossler, X.H. Yu,
   Phys. Rev. Lett., {\bf 62}, 2317 (1989).
\bibitem{BKT-IV} 
   A. K. Pradhan, S. J. Hazell, J. W. Hobdy, C. Chen, Y. Hu, B. M. Wanklyn,
   Phys. Rev. B {\bf 47}, 11374 (1993).
\bibitem{bib:3D-KT}
   M.B. Salamon, J. Shi, N. Overend, M.A. Howson, 
   Phys. Rev. B, {\bf 47}, 5520 (1993).\\
   S. Kamal, D.A. Bonn, N. Goldenfeld, P.J. Hirschfeld, R. Liang, W.N. Hardy,
   Phys. Rev. Lett., {\bf 73}, 1845 (1994).\\
   V. Pasler, P. Schweiss, C. Meingast, B. Obst, H. W${\rm \ddot u}$hl, A.I. Rykov, S. Tajima,
   Phys. Rev. Lett., {\bf 81}, 1094 (1998).\\
   K.D. Osborn, D.J. Van Harlingen, V. Aji, N. Goldenfeld, S. Oh, J.N. Eckstein,
   Phys, Rev, B, {\bf 68}, 144516 (2003).
\bibitem{bib:2D-KT}
   T. Schneider.
   EPL, {\bf 78}, 47003 (2007).
\bibitem{bib:Lemberger}
   I. Hetel, T. R. Lemberger, M. Randeria,
   Nature Physics, {\bf 3}, 700 (2007).\\
   Y.L. Zuev, J. A. Skinta, M. S. Kim, T. R. Lemberger, E. Wertz, K. Wu, Q. Li
   Physica C, 468, 276 (2008).
\bibitem{bib:BKT}
   V. L. Berezinskii,
   Zh. Eksp. Teor. Fiz. {\bf 61}, 1144 (1971)
   [Sov. Phys. JETP {\bf 34}, 610 (1972)].\\
   J. M. Kosterlitz, D. J. Thouless,
   J. Phys. C: Solid State Phys.
   {\bf 6}, 1181 (1973).
\bibitem{bib:Paramekanti}
   A. Paramekanti, M. Randeria, T. V. Ramakrishnan, S. S. Mandal,
   Phys. Rev. B., {\bf 62}, 6786 (2000).
\bibitem{bib:Kwon}
   H.J. Kwon, A.T. Dorsey, P.J. Hirschfeld,
   Phys. Rev. Lett., {\bf 86}, 3875 (2001).
\bibitem{bib:Franz}
   M. Franz, A.P. Iyengar,
   Phys. Rev. Lett., {\bf 96}, 047007 (2006).
   \bibitem{Herbut} I. F. Herbut and M. J. Case,
Phys. Rev. B {\bf 70}, 094516 (2004).

\bibitem{bib:MW}
   N. D. Mermin, H. Wagner,
   Phys. Rev. Lett. {\bf 17}, 1133 (1966). 
\bibitem{bib:Curty}
   P. Curty, H. Beck,
   Phys. Rev. Lett., {\bf 91}, 257002-1 (2003).

\bibitem{bib:Hufner}
   S. H$\rm{\ddot u}$fner, M.A. Hossain, A. Damascelli, G.A. Sawatzky,
   Rep. Prog. Phys. {\bf 71}, 062501 (2008).
\bibitem{bib:Panagopoulos99-2}
   C. Panagopoulos, R.J. Cooper, T. Xiang, Y.S. Wang, C.W. Chu,
   Phys. Rev. B {\bf 61}, R3808 (2000).
\bibitem{AA}
   E. Altman and A. Auerbach, 
   Phys. Rev. B {\bf 65}, 104508 (2002).
\bibitem{BFmodel} J. Ranninger, J.M. Robin, and M. Eschrig, Phys. Rev. Lett.
{\bf 74}, 4027 (1995)
\bibitem{QED3} M. Franz and Z. Tesanovic, Phys. Rev. Lett. {\bf 87}, 257003 (2001); M. Franz, Z. Tesanovic and O. Vafek,  Phys. Rev. B {\bf 66}, 054535 (2002); Z. Tesanovic, Nature Physics {\bf 4}, 408 (2008)

\bibitem{SubirBF} V. Galitski and S. Sachdev, Phys. Rev. B {\bf 79} 134512 (2009)

\bibitem{bib:Lee}
   W.S. Lee, I.M. Vishik, K. Tanaka, D.H. Lu, T. Sasagawa, N. Nagaosa, T.P. Devereaux, Z. Hussain, Z.X. Shen,
   Nature, {\bf 450}, 81 (2007).
   \bibitem{bib:Kanigel}   
   A. Kanigel, U. Chatterjee, M. Randeria, M. R. Norman, S. Souma, M. Shi, Z. Z. Li, H. Raffy, J. C. Campuzano,
   Phys. Rev. Lett. {\bf 99}, 157001 (2007).
\bibitem{bib:Hikami}
   S. Hikami, T. Tsuneto,
   Prog. Theor. Phys., {\bf 63}, 387 (1980).
\bibitem{IMFT}
   D.J. Scalapino, Y. Imry and P. Pincus, 
   Phys. Rev. B, {\bf 11}, 2042 (1975)\\  
   See also R. Ofer \etal, Phys. Rev. B {\bf 74}, 220508 (2006) for a more recent application.
   \bibitem{comm-core}
      We note that in the detailed analysis of Benfatto {\em et. al.} \cite{bib:Giamarchi}, the effect of a large vortex core energy was emphasized.
   They have found that it can produce a large numerical factor in front of Eq. (\ref{DeltaTc}).  Here we  choose, for simplicity,
   to depict the results for the standard $xy$ model keeping in mind that vortex core energy  effects might be important for future comparisons with experiments.

\bibitem{bib:Giamarchi}
   L. Benfatto, C. Castellani, T. Giamarchi
   Phys. Rev. Lett., {\bf 98}, 117008 (2007).
\bibitem{bib:Kosterlitz}
   J. M. Kosterlitz,
   J. Phys. C: Solid State Phys., {\bf 7}, 1046 (1974).
\bibitem{WenLee}, P. A. Lee and X.-G.Wen, Phys. Rev. Lett. {\bf 78}, 4111 (1997).
\bibitem{bib:Maki}
   H. Won, K. Maki,
   Phys. Rev. B. {\bf 49}, 1397 (1994).



\bibitem{bib:Farnan}
   G.A. Farnan, G.F. Cairns, P. Dawson, S.M. O'Prey, M.P. McCurry, D.G. Walmsley,
   Physica C, {\bf 403}, 67 (2004).
\bibitem{bib:Motohashi}
   T. Motohashi, J. Shimoyama, K. Kitazawa, K. Kishio, K.M. Kojima, S. Uchida, S. Tajima,
   Phys. Rev. B, {\bf 61}, R9269 (2000).
\bibitem{bib:Colson}
   S. Colson, C.J. van der Beek, M. Konczykowski, M.B. Gaifullin, Y. Matsuda, P. Gierlowski, M. Li, P.H. Kes,
   Physica C, {\bf 369}, 236 (2002).
\bibitem{bib:Dordevic05}
   S.V. Dordevic, Y.J. Wang, D.N. Basov,
   Phys. Rev. B, {\bf 71}, 054503 (2005).
\bibitem{bib:Ortolani}
   M. Ortolani, S. Lupi, V. Morano, P. Calvani, P. Masselli, L. Maritato, M. Fujita, K. Yamada, M. Colapietro,
   Journal of Superconductivity and Novel Magnetism, {\bf 17}, 127 (2004).
\bibitem{bib:Thorsmolle}
   V.K. Thorsm${\rm\not o}$lle, R.D. Averitt, M.P. Maley, L.N. Bulaevskii, C. Helm, A.J. Taylor,
   Optics Letters, {\bf 26}, 1292 (2001).
\bibitem{bib:Qiu}
   X.G. Qiu, H. Koinuma, M. Iwasaki, T. Itoh, A.K. Sarin Kumar, M. Kawasaki, E. Saitoh, Y. Tokura,  K. Takehana, G. Kido, Y. Segawa,
   Appl. Phys. Lett., {\bf 78}, 506 (2001).
\bibitem{bib:Kojima}
   K.M. Kojima, S. Uchida, Y. Fudamoto, S. Tajima,
   Physica C, {\bf 392}, 57 (2003).
\bibitem{bib:Uchida}
   S. Uchida, K. Tamasaku,
   Physica C, {\bf 293}, 1 (1997).
\bibitem{bib:Dordevic03}
   S.V. Dordevic, S. Komiya, Y. Ando, D.N. Basov,
   Phys. Rev. Lett., {\bf 91}, 167401-1 (2003).
\bibitem{bib:Tominari}
   Y. Tominari, T. Kiwa, H. Murakami, M. Tonouchi, H. Schneidewind,
   Appl. Phys. Lett., {\bf 80}, 3147 (2002).
\bibitem{bib:Kitano}
   H. Kitano, T. Hanaguri, Y. Tsuchiyda, K. Iwaya, R. Abiru, A. Maeda,
   J. Low Temp. Phys., {\bf 117}, 1241 (1999).
\bibitem{bib:Timusk}
   T. Timusk, B. Statt,
   Rep. Prog. Phys. {\bf 62}, 61 (1999).
\bibitem{bib:Panagopoulos98}
   C. Panagopoulos, J.R. Cooper, T. Xiang,
   Phys. Rev. B, {\bf 57}, 13422 (1998).
\bibitem{bib:Gaifullin}
   M.B. Gaifullin, y. Matsuda, N. Chikumoto, J. Shimoyama, K. Kishio,
   Phys. Rev. Lett., {\bf 84}, 2945 (2000).
\bibitem{bib:Panagopoulos99-1}
   C. Panagopoulos, B.D. Rainford, J.R. Cooper, W. Lo, J.L. Tallon, J.W. Loram, J. Betouras, Y.S. Wang, C.W. Chu,
   Phys. Rev. B. {\bf 60}, 14617 (1999).
\bibitem{bib:Wang}
   Y.T. Wang, A.M. Hermann,
   Physica C, {\bf 335}, 134 (2000).
\bibitem{comm-dark} 
   The far ('dark') sides of the small pockets have not yet been unambigously detected, perhaps due to small quasiparticle normalization factors. 
\bibitem{twogaps} T. Kondo,  T. Takeuchi, A. Kaminski,  S. Tsuda, and S. Shin,
Phys. Rev. Lett. {\bf 98}, 267004 (2007).

\bibitem{Raman}
   M. Le Tacon, A. Sacuto, A. Georges, G. Kotliar, Y. Gallais, D. Colson, A. Forget, 
   Nature Physics 
   {\bf 2}, 537 (2006).
\bibitem{Ginzburg}
   V.L. Ginzburg, 
   Fiz. Tverd. Tela {\bf 2}, 203 (1960) 
   [Sov. Phys.-Solid State {\bf 2}, 1824 (1960)].
\bibitem{CL} 
   P.M. Chaikin and T.C. Lubensky, 
   ''Principles of Condensed Matter Physics'',
   Cambridge University Press (1995), Ch.~5.1.
 
 
 \bibitem{bib:Tobochnik}
   Tobochnik J., Chester G. V.,
   Phys. Rev. B, {\bf 20}, 3761  (1979).
\bibitem{bib:T_KT}
   J. F. Fernandez, M. F. Ferreira, J. Stankiewicz,
   Phys. Rev. B. {\bf 34}, 292 (1986).\\
   R. Gupta, J. DeLapp, G. G. Batrouni, G. C. Fox, C. F. Baillie, J. Apostolakis,
   Phys. Rev. Lett. {\bf 61}, 1996 (1988).

\end{thebibliography}
\end{document}